\PassOptionsToClass{natbib=false}{acmart}
\documentclass[sigplan,10pt]{acmart}
\usepackage[style=acmnumeric]{biblatex}
\newcommand*{\citet}[1]{\textcite{#1}}

\usepackage[nameinlink, noabbrev, capitalise]{cleveref}
\crefformat{section}{\S#2#1#3}
\crefformat{subsection}{\S#2#1#3}
\crefformat{subsubsection}{\S#2#1#3}
\crefrangeformat{section}{\S#3#1#4--#5#2#6}
\crefmultiformat{section}{\S#2#1#3}{ and~\S#2#1#3}{, \S#2#1#3}{ and~\S#2#1#3}

\usepackage{mathpartir}
\usepackage{float}

\newcommand*{\ind}{Indirection Theory}

\newcommand*{\m}[1]{\mathit{#1}}

\renewcommand{\implies}{\Rightarrow}
\renewcommand{\iff}{\Leftrightarrow}
\newcommand*{\defeq}{\triangleq}
\newcommand*{\ap}{\mkern3mu}
\newcommand*{\finmapsto}{\rightharpoonup_{\mathrm{fin}}}
\newcommand*{\all}[1]{\forall\, #1.\,}
\newcommand*{\ex}[1]{\exists\, #1.\,}

\newcommand*{\Vrel}{\mathcal{V}}
\newcommand*{\Erel}{\mathcal{E}}
\newcommand*{\Hrel}{\mathcal{H}}
\newcommand*{\Krel}{\mathcal{K}}
\newcommand*{\Rrel}{\mathcal{R}}
\newcommand*{\Grel}{\mathcal{G}}
\newcommand*{\Lrel}{\mathcal{L}}
\newcommand*{\Orel}{\mathcal{O}}
\newcommand*{\Erelo}{\Erel^\circ}
\newcommand*{\heapsat}{\Hrel}
\newcommand*{\tieWF}{\m{tieWF}}
\newcommand*{\heapWF}{\m{heapWF}}

\newcommand*{\ehole}{\boxdot}
\newcommand*{\ealloc}[1]{\m{ref}\ap#1}
\newcommand*{\ederef}[1]{{!}#1}
\newcommand*{\eassign}[2]{#1 \mathbin{:=} #2}
\newcommand*{\efree}[1]{\m{free}\ap#1}
\newcommand*{\econfig}[2]{\langle #1,#2\rangle}
\newcommand*{\vlambda}[2]{\lambda #1.\, #2}

\newcommand*{\Prop}{\m{Prop}}
\newcommand*{\pure}[1]{\ulcorner #1 \urcorner}
\newcommand*{\pfalse}{\m{False}}
\newcommand*{\ptrue}{\m{True}}

\newcommand*{\iProp}{\m{iProp}}
\newcommand*{\uPred}{\m{uPred}}

\newcommand*{\iRes}{\m{iRes}}

\newcommand*{\gname}{\gamma}

\newcommand*{\mask}{\mathcal{E}}

\newcommand*{\GHeap}{\m{GHeap}}
\newcommand*{\gheap}{\gname_{heap}}
\newcommand*{\concheap}{\sigma}
\newcommand*{\stateinterp}{\m{SI}}
\newcommand*{\wsat}{\m{wsat}}
\newcommand*{\regAuth}{\m{regAuth}}
\newcommand*{\gregistry}{\gname_\m{reg}}
\newcommand*{\slot}{\m{invState}}
\newcommand*{\disabled}{\m{disabled}}
\newcommand*{\enabled}{\m{enabled}}
\newcommand*{\disjmask}{\m{disjMask}}
\newcommand*{\genabled}{\gname_{\m{en}}}
\newcommand*{\gdisabled}{\gname_{\m{dis}}}
\newcommand*{\DisjMask}{\m{DisjMask}}
\newcommand*{\topmask}{\top}

\newcommand*{\proves}{\vdash}

\newcommand*{\sep}{\mathbin{*}}
\newcommand*{\bigsep}{\mathop{\vcenter{\hbox{\scalebox{2}{$\ast$}}}}\displaylimits}
\newcommand*{\wand}{\mathrel{-\!\!*}}
\newcommand*{\later}{\mathop{\triangleright}}
\newcommand*{\pers}{\mathop{\square}}

\newcommand*{\exZero}{\mathop{\diamond}}
\newcommand*{\bupd}{\mathop{\mathord{\mid}\!\!\!\Rrightarrow}}

\newcommand*{\basecell}{\m{BaseCell}}
\newcommand*{\knowInv}[2]{\boxed{#2}^{\,#1}}
\newcommand*{\ownGhost}[2]{\m{own}\ap#1\ap#2}
\newcommand*{\ownw}{\m{Own}}
\newcommand*{\own}[1]{\ownw(#1)}

\newcommand*{\wpE}[3]{\m{wp}_{#1}\ #2\ \{#3\}}
\newcommand*{\hoare}[4]{\{#1\}\ #2\ \{#3\}_{#4}}
\newcommand*{\wpstate}{\m{Res}}

\newcommand*{\ndist}[1]{\mathrel{\equiv_{#1}}}

\newcommand*{\rop}{\mathbin{\cdot}}
\newcommand*{\unit}{\varepsilon}
\newcommand*{\core}[1]{\mathop{\m{core}}#1}
\newcommand*{\pcore}[1]{\mathop{\m{pcore}}#1}
\newcommand*{\some}[1]{\mathop{\m{some}}#1}
\newcommand*{\none}{\m{none}}
\newcommand*{\valid}[1]{\mathop{\checkmark}#1}
\newcommand*{\validN}[2]{\mathop{\checkmark_{#1}}#2}
\newcommand*{\incl}{\preccurlyeq}
\newcommand*{\inclN}[1]{\mathrel{\preccurlyeq_{#1}}}

\newcommand*{\Ag}{\m{Ag}}

\newcommand*{\authfull}[1]{\bullet\, #1}
\newcommand*{\authfrag}[1]{\circ\, #1}

\newcommand*{\RawPredAt}{\m{RawPredAt}}
\newcommand*{\Knot}{\m{Knot}}

\newcommand*{\HPred}{\m{HPred}}
\newcommand*{\RawPred}{\m{RawPred}}
\newcommand*{\Res}{\m{Res}}
\newcommand*{\rmap}{\m{map}}
\newcommand*{\ageOne}{\m{age}_1}
\newcommand*{\agestep}{\mathrel{\shortdownarrow}}
\newcommand*{\ageRT}{\mathrel{\agestep^{*}}}
\newcommand*{\ageT}{\mathrel{\agestep^{+}}}
\newcommand*{\ext}{\preccurlyeq}
\newcommand*{\mono}{\mathrel{\sqsubseteq}}
\newcommand*{\klvl}[1]{|#1|}
\newcommand*{\kres}[1]{\m{res}(#1)}
\newcommand*{\stratify}{\m{stratify}}
\newcommand*{\unstratify}{\m{unstratify}}

\newcommand*{\hered}{\m{hereditary}}

\usepackage{stmaryrd}

\setcopyright{acmlicensed}
\copyrightyear{2018}
\acmYear{2018}
\acmDOI{XXXXXXX.XXXXXXX}
\acmConference[Conference acronym 'XX]{Make sure to enter the correct
  conference title from your rights confirmation email}{June 03--05,
  2018}{Woodstock, NY}
\acmISBN{978-1-4503-XXXX-X/2018/06}

\begin{document}

\title[Iris on Indirection Theory: Lightweight Higher-order Ghost State]{Iris on Indirection Theory: \texorpdfstring{\\}{}Lightweight Higher-order Ghost State}

\author{Darius Foo}
\orcid{0000-0002-3279-5827}
\correspondingauthor
\affiliation{%
  \institution{National University of Singapore}
  \country{Singapore}
}
\email{dariusf@u.nus.edu}

\author{Viet-Anh Nguyen}
\affiliation{%
  \institution{National University of Singapore}
  \country{Singapore}
  }
\email{vanh10@nus.edu.sg}

\author{Wei-Ngan Chin}
\affiliation{%
  \institution{National University of Singapore}
  \country{Singapore}
  }
\email{chinwn@nus.edu.sg}

\renewcommand{\shortauthors}{Foo et al.}

\begin{abstract}
Today, semantic models of languages with concurrency and higher-order state are constructed using
step-indexed predicates over resources and higher-order ghost state.
There are two main lines of work for constructing such models.
On the one hand, Iris provides these in full generality,
at the cost of a large body of mechanised domain theory
that is difficult to learn and port to new proof assistants.
On the other hand, \ind{} is much simpler
through the use of ordinary dependent types,
but its foundation in separation algebra limits the expressiveness
of its resource model compared to Iris-style higher-order ghost state.
We resolve this conundrum by showing that \ind{} can be as expressive as Iris.
To that end, we give a lightweight mechanisation of \ind{} in Lean,
generalised so knots contain resources from an Iris-style resource algebra functor.
We then reconstruct the Iris base logic on top of it.
The advantage of this approach is in the elementary metatheory:
a user can follow the whole construction without domain theory.
We conduct two case studies to show that the base logic is expressive and faithful to Iris:
a biorthogonal logical relation for higher-order state,
and an Iris-style program logic with invariants and user-definable ghost state,
complete enough to verify concurrent and higher-order programs.
\end{abstract}




\maketitle

\section{Introduction}
\label{sec:intro}

In the semantics of programming languages, one often encounters definitions that are apparently circular, and hence not well-founded.
Step-indexing breaks this circularity by stratifying the definitions by a natural number \emph{index}.

For example, to define a (binary) logical relation for the untyped lambda calculus, we can state the relation on values $\Vrel$ as follows ($\Erel$ is the relation on expressions).
\begin{mathpar}
\begin{array}{l}
  \Vrel_n \defeq \{ (\vlambda{x}{e_1}, \vlambda{x}{e_2}) \mid \all{k < n} \\
  \quad \all{(v_1, v_2) \in \Vrel_k} (e_1[v_1/x], e_2[v_2/x]) \in \Erel_k \}
\end{array}
\end{mathpar}
Having the recursive occurrences at $k < n$ makes the definition well-founded.

Despite its elegance and conceptual simplicity,
explicit low-level index manipulation does not scale well,
obscuring proofs with index arithmetic.
Modern presentations thus hide indices behind modal logics~\cite{DBLP:conf/lics/DreyerAB09,DBLP:conf/popl/AppelMRV07},
which provide essential high-level facilities such as defining guarded recursive predicates,
and reasoning by L\"ob induction.
Lightweight implementations of these modal logics exist.
In IxFree~\cite{polesiuk2017ixfree}, for example, one can write the following definition instead,
using the $\later$ (later) modality.
\begin{mathpar}
\begin{array}{l}
  \Vrel \defeq \{ (\vlambda{x}{e_1}, \vlambda{x}{e_2}) \mid \\
  \quad \all{(v_1, v_2) \in \later \Vrel} (e_1[v_1/x], e_2[v_2/x]) \in \later \Erel \}
\end{array}
\end{mathpar}

For this calculus, the domain of step-indexed propositions that IxFree provides is sufficient.
For more advanced language features such as state and concurrency, however,
we would need a domain of \emph{propositions over heaps}, and even defining one is a problem.
A classic example of this occurs in a language with first-class locks~\cite{DBLP:conf/esop/HoborAN08},
each guarding a \emph{resource invariant}~\cite{DBLP:journals/tcs/OHearn07},
a proposition describing the resources the lock protects.
A naive model may be given as follows.
\begin{align*}
\m{Heap} & \defeq \m{Loc} \finmapsto \m{Res} \\
\m{Res} & \defeq \m{Val} \mid \m{Lock}\ap\m{HProp} \\
\m{HProp} & \defeq \m{Heap} \to \m{Prop}
\end{align*}
There is a circularity in the semantic domains: unfolding $\m{Heap}$ and $\m{HProp}$ in $\m{Res}$ gives
\[
\m{Res} \defeq \m{Val} \mid \m{Lock}\ap((\m{Loc} \finmapsto \m{Res}) \to \m{Prop})
\]
where $\m{Res}$ occurs negatively.
This equation has no solution in the category of sets, and a proof assistant would reject it.

The problem of circular domain equations has been extensively studied, and has
led to
two main lines of work.
The more well-known one is
Iris~\cite{DBLP:journals/jfp/JungKJBBD18},
which solves the equations in the category of \emph{complete ordered families of equivalences} (COFEs).
The other is \ind, which uses ``only simple, elementary mathematics''~\cite{DBLP:conf/popl/HoborDA10},
i.e., ordinary dependent types, to construct a \emph{knot} containing approximations of predicates.

Researchers building on these frameworks today face a tradeoff.
Due to its maturity and generality, Iris is a large and complex framework, and working at its core demands deep expertise.
The developers of PulseCore noted the requirement for ``considerable expertise and effort in mechanized domain theory''~\cite{ebner2025pulsecore} when considering a port to F$\star$.
A faithful port such as
Iris-Lean~\cite{konig2022improved,irislean} is a similarly substantial development of about 60K lines of code.
In contrast,
\ind{} is substantially simpler
(the original construction is only 400 LoC~\cite{DBLP:conf/popl/HoborDA10},
though that is the core alone, without a base logic, ghost state or a program
logic),
and has been used successfully in the Verified Software Toolchain (VST)~\cite{DBLP:conf/nfm/Appel12,msl} and more recently in PulseCore~\cite{ebner2025pulsecore}.
However, it is built on \emph{separation algebra},
which restricts the model from having the full expressiveness of Iris's higher-order ghost state:
\ind-based VST ``could express only limited higher-order ghost state'',
motivating a reimplementation on top of Iris~\cite{DBLP:journals/pacmpl/ManskyD24},
and PulseCore implements ``two primitive kinds of higher order ghost state: invariants and predicates'',
relying on encodings for the others~\cite{ebner2025pulsecore}.
Whether these limitations are inherent or an artifact of these developments is unclear.

We show that the limitations are not inherent.
To that end, we develop a \emph{lightweight} mechanisation of \ind{} in Lean~\cite{DBLP:conf/cade/Moura021},
generalised to carry resources from a \emph{resource algebra functor}.
It is lightweight in that the metatheory is elementary: a user can follow the whole construction without domain theory.
We implement the Iris \emph{base logic} on top of it, which is small yet expressive.
The development is self-contained and compact, taking under 1,700 LoC.

To demonstrate that the base logic is faithful to Iris,
we use it in two case studies:
a biorthogonal logical relation for a heap-manipulating language,
and a program logic for a concurrent, heap-manipulating one.
The program logic is complete enough to verify concurrent and higher-order programs,
as we show on several small examples.

In summary, we make the following contributions:

\begin{itemize}
  \item A lightweight implementation of \ind{} in Lean, generalised to support user-definable higher-order ghost state, thus making the construction of higher-order semantic models accessible to a wider audience and clarifying the connection to Iris's machinery (\cref{sec:library})
  \item A case study in direct use of the library: a biorthogonal logical relation for a language with higher-order state (\cref{sec:logrel})
  \item A reconstruction of Iris-style higher-order ghost state, impredicative invariants, and weakest preconditions atop \ind{}, complete enough to verify concurrent and higher-order programs (\cref{sec:iris})
\end{itemize}

\section{Background}
\label{sec:background}

In this section, we recall the parts of Iris that our development reconstructs.
Iris is a higher-order concurrent separation logic.
Reasoning about programs with it is carried out using a program logic with advanced features, such as \emph{invariants} and \emph{fancy updates}.
These features are not primitive but derived~\cite{DBLP:conf/esop/Krebbers0BJDB17} from a much smaller base logic, whose propositions have type $\iProp$ (\cref{sec:bg:base_logic}).
The base logic is in turn an instance of a more general base logic of resources, whose propositions have type $\uPred$ (\cref{sec:bg:upred_to_iprop}).
Constructing $\iProp$ and $\uPred$ requires Iris's advanced structures (\cref{sec:bg:cofes_and_ucmras});
our aim is to construct the corresponding version of $\iProp$ through simpler structures (\cref{sec:bg:target}).

\begin{figure}
  \small
  \[
    \begin{array}{@{}r@{\ }l@{\quad}l@{}}
      P, Q \mathrel{::=}{}
        & \pure{\phi} \mid P \wedge Q \mid P \vee Q \mid P \implies Q \\
      \mid{} & \all{x} P \mid \ex{x} P
        & \\
      \mid{} & P \sep Q \mid P \wand Q
        & \text{\footnotesize(separation)} \\
      \mid{} & \later P \mid t \ndist{} u
        & \text{\footnotesize(step-indexing)} \\
      \mid{} & \pers P \mid \own{a} \mid \bupd P
        & \text{\footnotesize(resources)}
    \end{array}
  \]
  \Description{The primitive connectives of the Iris base logic.}
  \caption{The primitive connectives of the Iris base logic.}
  \label{fig:iris_base_logic}
\end{figure}

\subsection{The base logic \texorpdfstring{$\iProp$}{iProp}}
\label{sec:bg:base_logic}

The Iris base logic~\cite{DBLP:journals/jfp/JungKJBBD18} is a higher-order affine separation logic with two further groups of connectives (\cref{fig:iris_base_logic}).
The first has to do with step-indexing: the later modality $\later$ and an
internalised equality $t \ndist{} u$.
The second has to do with separation logic resources.
$\own{a}$ asserts ownership of the resource $a$.
The persistence modality $\pers P$ says that $P$ holds of the \emph{core} of the resource at hand, a duplicable part of it.
The \emph{basic update} $\bupd P$ says the resource at hand can be updated to one satisfying $P$, without invalidating any \emph{framing} resource.
We make this more precise in \cref{sec:bg:cofes_and_ucmras}.
Invariants, higher-order ghost state and fancy updates are derived from these.

\subsection{From \texorpdfstring{$\uPred$}{uPred} to \texorpdfstring{$\iProp$}{iProp}}
\label{sec:bg:upred_to_iprop}

To see how $\iProp$ is constructed, we begin with
\emph{uniform predicates} $\uPred$,
predicates on a step index and a type of resource $M$.
\[
  \begin{array}{@{}l@{}}
    \uPred(M) \defeq \{ P : \mathbb{N} \to M \to \Prop \mid \all{n_1, n_2, x_1, x_2} \\
    \quad n_2 \le n_1 \implies x_1 \inclN{n_2} x_2 \implies P\ap n_1\ap x_1 \implies P\ap n_2\ap x_2 \}
  \end{array}
\]
Intuitively, uniform predicates remain true (1) when there are fewer steps left to observe, and (2) on larger resources.
The first condition makes them \emph{downward-closed} in the step index:
a $\uPred$ remains true one step later, i.e. $P \proves \later P$.
The second condition makes them \emph{upward-closed} under resource extension $\inclN{n}$,
in turn making them affine, i.e. $P \sep Q \proves P$.

To reach $\iProp$, we need to instantiate $\uPred$ with some type of resource $M$, which we have thus far left abstract.
As a first approximation, we give three requirements on $M$.

Firstly, it must have enough structure to interpret the connectives of \cref{fig:iris_base_logic}:
composition $\cdot$ to split resources for $\sep$,
a duplicable \emph{core} for $\pers$, and a notion of when a resource is valid to ensure $\bupd$ does not invalidate frames.
Iris packages these properties as the notion of a \emph{unital camera};
more detail on this is given in \cref{sec:bg:cofes_and_ucmras}.
Secondly, $M$ must be able to contain propositions $\iProp$, which we have not yet defined.
Therefore, what we instantiate $M$ with must be defined mutually recursively with $\iProp$.

Given $M$ is a user-chosen
\emph{unital camera functor} $F$,
Iris takes the fixed point of the following equations.
\[
  \iRes \cong F(\iProp)
  \qquad
  \iProp \defeq \uPred(\iRes)
\]
Unfolding, this is $\iRes \cong F(\uPred(\iRes))$, a recursive domain equation.
Its solution gives us $\iProp$.

\subsection{Constructing \texorpdfstring{$\uPred$}{uPred}: COFEs and UCMRAs}
\label{sec:bg:cofes_and_ucmras}

Making \cref{sec:bg:upred_to_iprop} precise requires a substantial amount of structure,
almost all of it for solving the domain equation.
We recall it briefly, to fix terminology, and because \cref{sec:lib:comparison_with_iris} compares it against what \ind{} needs.

Because of cardinality, no type satisfies the equation as given in the category of sets.
The idea of the solver is to move to a setting of types with approximate equality,
under which two elements are considered equal up to $n$ steps if they cannot be distinguished within
that many
steps.
A solution is then a type isomorphic to its own unfolding with respect to approximate equality.
An OFE provides a canonical structure for capturing this notion of approximation. Cameras then equip resources with an OFE structure. The OFE structure serves purposes beyond solving the recursive domain equations: in particular, the stronger COFE structure supports L\"ob induction and guarded recursive definitions.

\paragraph{(C)OFEs.}
An \emph{ordered family of equivalences} is a type $A$ with an equivalence $\equiv$ and a family of equivalences $(\ndist{n})_{n \in \mathbb{N}}$, satisfying
$x \equiv y \Leftrightarrow \all{n} x \ndist{n} y$
and
$n \le m \implies x \ndist{m} y \implies x \ndist{n} y$.
Every type carrying step-indexed data must be an OFE, and every function between such types must be \emph{non-expansive}: $x \ndist{n} y \implies f(x) \ndist{n} f(y)$.
It is \emph{contractive} if the weaker $\all{m < n} x \ndist{m} y$ already gives $f(x) \ndist{n} f(y)$.
An OFE may further be \emph{complete} (a COFE), which gives contractive functions unique fixed points, allowing guarded recursive definitions.

\paragraph{Resource algebras and (unital) cameras}
A type of resources satisfying the requirements of \cref{sec:bg:upred_to_iprop} is called a \emph{resource algebra} (RA).
Concretely, it is a type with a composition operation $\rop$ which combines two resources, a predicate $\valid{}$ which says whether a resource is valid, and a partial \emph{core} operation $\pcore{}$ which returns the duplicable part of a resource when it has one.
A \emph{camera} (CMRA) is a resource algebra whose carrier is an OFE, so that its resources inherit the corresponding notion of approximate equality. In addition, a camera is equipped with step-indexed validity $\valid{}_n$, which can be viewed as the validity analogue of step-indexed equality.
A \emph{unital} camera (UCMRA) has an additional empty resource $\unit$.

\paragraph{Functors and the solver.}
Since resources store propositions, the user does not supply a unital camera but a \emph{unital camera functor} $F$: given a COFE of propositions, $F$ builds the unital camera of resources storing them. $F$ must be \emph{locally contractive}, meaning that resources built from propositions equal up to $n$ steps are equal up to $n+1$ steps. The \emph{solver} is the theorem that such an $F$ admits the fixed point of \cref{sec:bg:upred_to_iprop}, a COFE $X$ with $X \cong \uPred(F(X))$.

\paragraph{L\"ob induction and guarded recursive definition.} In Iris, $\iProp$ is a COFE, which allows us to reason by Löb induction; under the hood, this corresponds to strong induction on the step-index. $\iProp$ also supports guarded recursive definitions. In such definitions, recursive occurrences typically appear under the $\later$ modality, which ensures that they are contractive and hence admit a unique fixed point.

\subsection{An alternative route to \texorpdfstring{$\iProp$}{iProp}}
\label{sec:bg:target}

The Iris base logic (\cref{sec:bg:base_logic,fig:iris_base_logic}) is expressive, well-understood, and widely used.
We thus take
$\iProp$ as a specification,
viewing the construction of \cref{sec:bg:upred_to_iprop,sec:bg:cofes_and_ucmras} as one way to satisfy it.
We then explore the use of \ind{}~\cite{DBLP:conf/popl/HoborDA10} as an alternative way to realise this base logic,
in a form that suffices for common applications such as logical relations (\cref{sec:logrel}) and program logics (\cref{sec:iris}).
The next section (\cref{sec:library}) gives the construction, concluding with a comparison to Iris (\cref{sec:lib:comparison_with_iris}).

\section{The Indirection Theory Library}
\label{sec:library}

In this section, we construct a replacement for the base logic
of
\cref{fig:iris_base_logic}:
the type of propositions $\HPred(M)$,
standing for \emph{hereditary predicates},
over a user-chosen type of resources
$M$.
This is done in four layers:
resources (\cref{sec:lib:resources}),
knots (\cref{sec:lib:knots}),
hereditary predicates (\cref{sec:lib:hered}),
and logical connectives (\cref{sec:lib:base_logic}).
\cref{sec:lib:guarded_rec} then discusses two important features of
our step-indexing model: Löb induction and guarded recursive definitions.
Finally, \cref{sec:lib:comparison_with_iris} concludes the section by
comparing our construction with Iris's.

\subsection{Resources}
\label{sec:lib:resources}

The logic we are building is a separation logic.
We thus begin by recalling the notion of a resource algebra,
an abstraction of the resources the propositions of our logic describe.
An RA is a type $\alpha$ of resources equipped with a composition operator $\rop : \alpha \to \alpha \to \alpha$, a validity predicate $\valid{} : \alpha \to \Prop$, and an operation $\pcore{} : \alpha \to \m{Option}\ap\alpha$ (``partial core''), satisfying the laws of \cref{fig:ra_laws}.
Resource algebra \emph{inclusion} is defined as $a \incl b \defeq \ex{c} b = a \rop c$.

\begin{figure}[H]
  \small
    $\begin{array}{@{}l@{}}
      (a \rop b) \rop c = a \rop (b \rop c) \\
      a \rop b = b \rop a \\
      \valid{(a \rop b)} \implies \valid{a} \\
      \pcore{a} = \some{c} \implies c \rop a = a \\
      \pcore{a} = \some{c} \implies \pcore{c} = \some{c} \\
      a \incl b \implies \pcore{a} = \some{c_a} \implies \\[2pt] \quad
      \ex{c_b} \pcore{b} = \some{c_b} \wedge c_a \incl c_b
    \end{array}$
  \Description{}
  \caption{Resource algebra laws.}
  \label{fig:ra_laws}
\end{figure}

Composition is associative and commutative.
Validity is closed ``downward'', so subparts of a valid resource are also valid.
Some resources may be duplicated and shared freely:
$a$ is \emph{duplicable} if $a \rop a = a$.
$\pcore{a}$ is the duplicable part of $a$, if it exists.
The fourth law says $a$ contains its core,
and the fifth that the core is its own core.
Both of these together imply that the core is duplicable.
The last law says that the core is monotone:
if $a$ is included in $b$, the core of $a$ is included in the core of $b$.

A separation logic also needs a resource standing for ownership of nothing.
A \emph{unital resource algebra} (URA) is a resource algebra with such an element,
$\unit : \alpha$. The unit is valid ($\valid{\unit}$),
is the identity under composition ($\unit \rop a = a$),
and is its own core.
The unit is included in every resource,
so by monotonicity %
every element of a URA has a core, which we denote by $\core{a}$.

In a separation logic with higher-order ghost state,
a resource may contain a proposition of the logic, which itself describes a resource.
To break the cycle, we must abstract over the type $P$ of propositions.
Later, we will tie the knot by choosing $P$ to be the type of propositions being defined.

Next,
it is not enough to only parameterise resources over $P$,
as we will need to work with \emph{approximations} of the propositions stored in them.
We must be able to weaken these approximations without changing the structure of a resource.
This leads us to the (new) notion of a \emph{resource algebra functor} (RAF), a type $M : \m{Type} \to \m{Type}$.
$M(P)$ is the type of resources containing propositions $P$, and has an operation
$\rmap : (P \to Q) \to M(P) \to M(Q)$ satisfying the laws of
\cref{fig:ra_lawsf},
which are that
$\rmap$ is functorial (first two) and
commutes with all the resource algebra operations (next three).

\begin{figure}[H]
  \small
  \[
    \begin{array}{@{}l@{}}
      \rmap\ap\m{id}\ap a = a \\
      \rmap\ap(g \circ f)\ap a = \rmap\ap g\ap(\rmap\ap f\ap a) \\[3pt]
      \rmap\ap f\ap(a \rop b) = \rmap\ap f\ap a \rop \rmap\ap f\ap b \\
      \valid{a} \implies \valid{(\rmap\ap f\ap a)} \\
      \pcore{(\rmap\ap f\ap a)} = \m{Option.map}\ap(\rmap\ap f)\ap(\pcore{a}) \\[3pt]
      \rmap\ap f\ap a = x \rop y \implies \valid{(x \rop y)} \implies \\
      \quad \ex{x'\,y'} a = x' \rop y' \wedge \rmap\ap f\ap x' = x \wedge \rmap\ap f\ap y' = y
    \end{array}
  \]
  \Description{}
  \caption{Resource algebra functor laws.}
  \label{fig:ra_lawsf}
\end{figure}

A \emph{unital} resource algebra functor is simply a resource algebra functor $M$
where $M(P)$ is a unital resource algebra.
The last law of \cref{fig:ra_lawsf} lets us derive that
map preserves the unit, i.e. $\rmap\ap f\ap\unit = \unit$, which would otherwise
have to be assumed.

\subsection{Knots}
\label{sec:lib:knots}

We now have
resources that store propositions of an arbitrary type $P$, but
not yet the type of propositions itself.

Propositions should be predicates on
resources, and resources should store propositions, so the type $P$ we want
would have to satisfy $P \cong M(P) \to \Prop$.
This has no solution in set theory.
Iris solves
a variant of
this equation in the category of COFEs.
We will
take a more elementary route,
based on the
idea of working with \emph{approximations} of propositions.

The idea is to assign every proposition a \emph{level} $n \in \mathbb{N}$.
A proposition of level $n{+}1$ is a predicate on resources storing propositions of level $n$.
At level $0$, there are no lower levels to refer to, so there is a single
trivial proposition.
Intuitively, the level says how good an approximation a proposition is, or how much information it has.
More concretely, it measures how deeply a proposition can inspect the propositions stored in the resources it describes.

One more refinement is required.
A proposition true at level $n{+}1$ must remain true at every lower level, as it contains ``more'' information.
However, a predicate on level-$n$ resources says nothing about resources at lower levels.
We therefore pair
each predicate with its own ``truncation'' to the level below, so that a
level-$n$ proposition is really a sequence of $n$ predicates
of successively smaller levels.

We now make this precise, building up to the propositions of the logic (\cref{sec:lib:hered}).
Let $M$ be a (user-chosen) resource algebra functor.
We write $\Res(P) \defeq \{a : M(P) \mid \valid{a}\}$ for a valid resource of $M$.
We then define a type of ``raw'' predicates \emph{at} a level:
\[
  \begin{aligned}
    \RawPredAt_0 & \defeq \m{unit} \\
    \RawPredAt_{n+1} & \defeq \RawPredAt_n \times (\Res(\RawPredAt_n) \to \Prop)
  \end{aligned}
\]
The first element of the pair in the recursive case is the ``truncation'' of the predicate,
while the second element is the predicate itself.
Thus these predicates are over resources containing the (raw) predicates of the level below,
the first step towards tying the knot as promised earlier.

$\RawPredAt$ is a type family, i.e. with one type per level, but we would like a type of propositions usable at every level.
The next step towards that is to hide the level using a dependent pair.
A \emph{knot} is hence a level and a valid resource of that level.
\[
  \Knot \defeq \Sigma\,(n : \mathbb{N}).\ \Res(\RawPredAt_n)
\]
For a knot $k$, we write $\klvl{k}$ for its level $n$ and $\kres{k}$ for its resource.
We also name the type of predicates over knots.
\[
  \RawPred \defeq \Knot \to \Prop
\]
The two types are related by a projection and an embedding.
$\stratify_n$ projects a predicate over knots onto a level-$n$ one, retaining its value at every knot $k$ with $\klvl{k} < n$.
$\unstratify_n$ embeds a level-$n$ predicate as a predicate over knots that is false at levels $\ge n$.
The two operations satisfy
\[
  \begin{array}{@{}l@{}}
    \stratify_n \circ \unstratify_n = \m{id} \\
    \klvl{k} < n \implies \unstratify_n(\stratify_n\ap P)\ap k = P\ap k
  \end{array}
\]
Differences from the original construction~\cite{DBLP:conf/popl/HoborDA10} are in the pervasive use of RA validity, and the use of a concrete knot and stratify/unstratify, rather than the abstract squash/unsquash.

\subsection{Hereditary predicates}
\label{sec:lib:hered}

A proposition true of a knot should remain true as the knot is approximated further or gains resources.
We laid some groundwork for the former earlier in $\RawPredAt$, but nothing yet ensures that a predicate over knots ($\RawPred$) will respect either.
We now define these two ways in which knots may change.
The predicates closed under them, \emph{hereditary predicates}, will be the propositions of the logic.

\paragraph{Aging}
The first way a knot changes is by \emph{aging},
which approximates it one level further.
A proposition that holds of a knot should continue to hold once the knot is aged, since an aged knot carries strictly less information.
Because the level strictly decreases, recursion through stored propositions is well-founded, which is what gives the logic $\later$ and L\"ob induction.

Aging a knot from level $n{+}1$ to $n$ replaces every stored proposition by its truncation.
Truncation of a $\RawPredAt_{n{+}1}$ simply projects its first component ($\pi_1$).
Because a knot contains a resource algebra functor, we can use its $\rmap$ to age all its stored propositions.
We can therefore define the function $\ageOne : \Knot \to \m{Option}\ap \Knot$ as follows.
\[
  \ageOne\ap(n{+}1, \phi) \defeq \some{(n, \rmap\ap\pi_1\ap\phi)}
  \qquad
  \ageOne\ap(0, \phi) \defeq \none{}
\]
We write $k \agestep k'$ to mean ``$k$ aged by one level is $k'$'', i.e. $\ageOne\ap k = \some{k'}$.
$\ageRT$ and $\ageT$ are its reflexive-transitive and transitive closures.
$\ageT$ strictly decreases the level, while $\ageRT$ never increases it.

\paragraph{Resource extension}
The second way a knot changes is by gaining resources.
A proposition that holds of a resource should continue to hold of a larger resource, which is what makes the logic affine and validates the frame rule.

Concretely, a knot may gain a frame at its own level.
\[
  k \ext k' \defeq \klvl{k} = \klvl{k'} \wedge \kres{k} \incl \kres{k'}
\]
Composing the two gives us a monotonicity relation:
\[
  k \mono k' \defeq \ex{k''} k \ageRT k'' \wedge k'' \ext k'
\]
\paragraph{Hereditary}
A predicate is \emph{hereditary} when it is closed under $\mono$.
\[
  \begin{array}{@{}l@{}}
    \hered\ap p \defeq \all{k\, k'} p\ap k \implies k \mono k' \implies p\ap k' \\
    \HPred \defeq \{p : \RawPred \mid \hered\ap p\}
  \end{array}
\]

$\ext$ was not part of the original construction~\cite{DBLP:conf/popl/HoborDA10}, but is present as the join order
$\le$
in MSL~\cite{msl}.
$\mono$ is Iris's $\uPred$ monotonicity condition.

\subsection{The base logic}
\label{sec:lib:base_logic}

Having defined knots and hereditary predicates,
we can finally define the connectives of the base logic (\cref{fig:base_logic_connectives}).
\begin{figure}[H]
  \small
  \[
    \begin{array}{@{}l@{\ }c@{\ }l@{}}
      P \proves Q & \defeq & \all{k} P\ap k \implies Q\ap k \\[2pt]
      (P \wedge Q)\ap k & \defeq & P\ap k \wedge Q\ap k \qquad (\all{x} P\ap x)\ap k \defeq \all{x} P\ap x\ap k \\
      (P \vee Q)\ap k & \defeq & P\ap k \vee Q\ap k \qquad (\ex{x} P\ap x)\ap k \defeq \ex{x} P\ap x\ap k \\
      (P \implies Q)\ap k & \defeq & \all{k'} k \mono k' \implies P\ap k' \implies Q\ap k' \\
      (P \sep Q)\ap k & \defeq & \ex{x_1\, x_2} \kres{k} = x_1 \rop x_2 \wedge P\ap(\klvl{k}, x_1) \wedge Q\ap(\klvl{k}, x_2) \\
      (P \wand Q)\ap k & \defeq & \all{k'\, z} k \ageRT k' \implies \valid{(\kres{k'} \rop z)} \implies \\
        & & \quad P\ap(\klvl{k'}, z) \implies Q\ap(\klvl{k'}, \kres{k'} \rop z) \\[2pt]
      (\bupd P)\ap k & \defeq & \all{k'\, z} k \ageRT k' \implies \valid{(\kres{k'} \rop z)} \implies \\
        & & \quad \ex{x} \valid{(x \rop z)} \wedge P\ap(\klvl{k'}, x) \\[2pt]
      (\pers P)\ap k & \defeq & P\ap(\klvl{k}, \core{(\kres{k})}) \\[2pt]
      (\later P)\ap k & \defeq & \all{k'} k \ageT k' \implies P\ap k' \\[2pt]
      (P \ndist{} Q)\ap k & \defeq & \all{k'} \klvl{k'} \le \klvl{k} \implies (P\ap k' \Leftrightarrow Q\ap k') \\[2pt]
      \own{a}\ap k & \defeq & a_{\klvl{k}} \incl \kres{k}
    \end{array}
  \]
  \Description{}
  \caption{Definition of the connectives of the base logic.}
  \label{fig:base_logic_connectives}
\end{figure}
Entailment, conjunction, disjunction, and quantifiers are pointwise.
Implication quantifies over the $\mono$-successors of $k$,
which is what makes it hereditary.
Separating conjunction splits the resource at the knot's own level.
The magic wand quantifies over $\ageRT$-successors and over frames, again to be hereditary.
The basic update modality is the primitive for updating ghost resources in the presence of a frame.
Persistence evaluates its argument at the core of the resource.
Later evaluates its argument at every knot strictly below the current level.
Internal equality of propositions holds at $k$ when both agree at every knot at or below $k$'s level.
The except-$0$ modality $\exZero P \defeq \later \pfalse \vee P$, meaning ``$P$ at every level above $0$'', is derived, and is used in \cref{sec:iris}.

Ownership is the only construct that differs significantly from Iris:
Iris's $\own{a}$ takes a fixed resource $a$,
but here the resource depends on the level, so a single $a$ does not suffice.
Here, $\ownw$ takes instead a \emph{coherent family} of resources $a : \all{n} M(\RawPredAt_n)$:
one resource per level, with a coherence condition saying that it is compatible with aging:
\[
  \rmap\ap\pi_1\ap(a_{n+1}) = a_n
\]
This is a sufficient condition to prove that $\ownw$ is hereditary.
The definition in \cref{fig:base_logic_connectives} reads: $\own{a}$ holds at a knot $k$ when the resource at $k$'s level, $a_{\klvl{k}}$, is included in $\kres{k}$.

We also overload the notation $\own{a}$ for the case where $a$ is an element of a resource algebra $A$, rather than a coherent family.
A resource algebra $A$ can be turned into a resource algebra functor using the \emph{constant} resource algebra functor
\[
  C(A)(P) \defeq A
\]
whose $\rmap$ ignores its first argument.
Coherence is then trivial by the definition of $C$.
We thus write $\own{a}$ for $\own{\lambda n.\, a}$ when $a$ comes from a resource algebra embedded via a constant functor; \cref{sec:logrel,sec:iris} both rely on this for the resources that do not depend on the level.

\subsection{L\"ob induction and guarded recursion}
\label{sec:lib:guarded_rec}
The logic of hereditary predicates enjoys
L\"ob induction, strong induction on the level of the knot,
as well as guarded recursion.
Definitions such as the weakest precondition of \cref{sec:iris} are recursive, and are taken as the fixed point of a contractive function $f : (A \to \HPred) \to (A \to \HPred)$ on families of propositions indexed by a type $A$ (the arguments of the recursive definition).
\[
  \begin{array}{@{}l@{}}
    \m{contractive}\ap f \defeq \all{P, Q, k} \\
    \quad(\all{x, k'} \klvl{k'} < \klvl{k} \implies
    (P\ap x\ap k' \Leftrightarrow Q\ap x\ap k')) \implies \\
    \quad\quad \all{x} f\ap P\ap x\ap k \Leftrightarrow f\ap Q\ap x\ap k
  \end{array}
\]
This intuitively means that recursive occurrences of the family being defined are guarded by $\later$.
Every contractive $f$ has a fixed point, $\m{Rec}\ap f = f(\m{Rec}\ap f)$, obtained by recursion on the level of the knot.

\subsection{Comparison with Iris}
\label{sec:lib:comparison_with_iris}

The differences between our construction and Iris's have been discussed throughout \cref{sec:background} and this section.
We recap them here.

\paragraph{Solving the domain equation}
Iris defines $\uPred(M)$ for any unital camera $M$, and obtains $\iProp$ by solving $\iRes \cong F(\uPred(\iRes))$ (\cref{sec:bg:cofes_and_ucmras}).
Requiring a single type isomorphic to its own unfolding is what necessitates the category of COFEs and the solver of~\citet{DBLP:journals/jcss/AmericaR89}.
We define $\HPred$ for any unital resource algebra functor $M$, and obtain the resources of each level $n$ by recursion on $\mathbb{N}$, storing the propositions of level $n{-}1$.
Requiring a type per level is what allows us to get away with an ordinary recursive definition, without the solver, at the cost of the level being part of every knot.

\paragraph{Propositions}
Iris propositions are predicates on $\mathbb{N} \times M$, so the index and the resource are independent.
Lowering the index leaves the resource unchanged.
Ours are predicates on $\Knot$, so the resource depends on the level.
Lowering the level maps the resource by $\rmap\ap\pi_1$.

\paragraph{Resources}
The user of Iris supplies a unital camera functor, so resources are OFEs and have step-indexed equality $\ndist{n}$ and validity $\validN{n}{}$.
The user of our logic supplies a unital resource algebra functor (\cref{sec:lib:resources}) instead, so resources only have ordinary equality and a single validity predicate.

\paragraph{Ownership}
Iris's $\own{a}$ takes a single $a : \iRes$, since the type of a resource does not depend on the index.
The same $a$ is used at every index.
Ours takes a coherent family (\cref{sec:lib:base_logic}), since the type of a resource depends on the level.
The resource $a_n$ is used at level $n$.

\paragraph{Internal equality}
Iris's internal equality works on both propositions and resources, as they are both OFEs.
Ours works on propositions only, with first-order resources handled with Leibniz equality.

\section{Case Study: Logical Relations}
\label{sec:logrel}

Our first case study uses the library directly, without the program logic of
\cref{sec:iris}.
We formalise a binary logical relation for an untyped,
call-by-value, \emph{stateful} lambda calculus,
extending
the biorthogonal
construction
of \citet{untypedlogicalrelation} with a heap.
\[
\begin{array}{@{}r@{\;}c@{\;}l@{}}
  e &::=& x \mid v \mid e\ap e \mid \ealloc{e} \mid \ederef{e} \mid \eassign{e}{e} \mid \efree{e} \\
  v &::=& \vlambda{x}{e} \mid \ell \\
  K &::=& \ehole \mid K\ap e \mid v\ap K \mid \ealloc{K} \mid
    \ederef{K} \mid \eassign{K}{e} \mid \eassign{v}{K} \mid \efree{K}
\end{array}
\]

A heap $h \in \m{Loc}\finmapsto\m{Option}\ap v$
is a finite, partial mapping of \emph{locations} to \emph{live cells} or \emph{tombstones}
(which represent deallocated locations).
For a location $\ell\in\operatorname{dom}(h)$, $h(\ell)=\none{}$ is a tombstone and $h(\ell)=\some{v}$ is a live cell holding $v$.
Allocation deterministically chooses a fresh location,
dereferencing and updating require a live cell,
and deallocation replaces a live cell with a tombstone.
Configurations $c ::= \econfig{h}{e}$ are pairs of a heap and a closed expression.
Evaluation contexts $K$ specify left-to-right call-by-value evaluation.
We write $K[e]$ for the expression obtained by filling the hole of $K$ with $e$.

\subsection{Biorthogonal logical relations}
A binary logical relation is a useful device for proving \emph{contextual equivalence},
a strong form of program equivalence under which no enclosing program context can distinguish two expressions by their observable behaviour.
The observable behaviour we use here is \emph{termination}: both expressions either terminate
or do not terminate.
To capture this notion,
we define an \emph{approximation} relation over two configurations $\Lrel(c_1,c_2)$ as follows.
If the expression in $c_1$ is already a value, then the expression in $c_2$ must terminate.
If $c_1$ instead takes a step to $c'_1$, $\Lrel(c'_1,c_2)$ must hold later.
The \emph{observation} relation $\Orel$ is $\Lrel$ in both directions.
\[
\begin{array}{r@{\,}l}
  \Lrel(c_1,c_2) \defeq &
    \m{value}(c_1) \implies \m{terminates}(c_2) \\
  & \wedge\ \all{c'_1} c_1 \longrightarrow c'_1 \implies
      \bupd\later \Lrel(c'_1,c_2) \\[2pt]
  \Orel(c_1,c_2) \defeq & \Lrel(c_1,c_2) \wedge \Lrel(c_2,c_1)
\end{array}
\]

The recursive occurrence of $\Lrel$ is guarded by $\later$, so its body is contractive,
allowing $\Lrel$ to be defined using the library's fixed-point operator (\cref{sec:lib:guarded_rec}).
The basic update permits ghost state to be updated between operational steps.
We use ghost state to define the notion of \emph{related heaps} next.

\subsection{Related heaps}
To relate two heap-manipulating programs, we must relate the two heaps they execute on.
Intuitively, the locations in the two heaps should correspond,
so we maintain a finite, partial bijection $T$
from locations in the left heap to locations in the right.

We record this bijection using the \emph{authoritative map} resource algebra,
which supports reasoning about a shared map.
One owner may hold the complete map as the authoritative source of truth,
while others hold \emph{fragments}: knowledge of specific entries in the map.
Fragments can be duplicated and distributed freely, but they must all remain compatible
with the authoritative map;
an update of the authoritative map must therefore preserve every fragment.
\begin{mathpar}
  \m{AuthMap}(A)
  \defeq
\m{Option}\bigl(\m{Ex}(\m{Loc}\finmapsto A)\bigr)
  \times (\m{Loc}\finmapsto A)

  \authfull{f} \defeq (\some{(\m{ex}(f))},\emptyset)

  \authfrag{g} \defeq (\none{},g)
\end{mathpar}
It is represented as a pair of an \emph{exclusive}, optional map and
a part of that map.
Exclusivity ensures that there can be at most one authoritative map at a time.
An element of this resource algebra is valid if the fragment component is included in the authoritative component, whenever the authoritative component is present.
$\authfull{f}$ and $\authfrag{g}$ denote full and fragmentary ownership of $f$ and $g$ respectively.

To allow several owners to agree on a single value, we use the \emph{agreement} resource algebra $\Ag(A)$.
Any number of owners may hold $\m{ag}(a)$ for a value $a$ and freely duplicate it,
but two such instances are valid only if they refer to the same value.
\[
  \valid{(\m{ag}(a) \rop \m{ag}(b))} \iff a = b
\]

The bijection is tracked using an authoritative map of agreements of locations $\m{AuthMap}(\Ag(\m{Loc}))$.
$\m{tieAuth}(T)$ asserts ownership of the authoritative finite map recording the mapping,
and fragment assertions $\m{tied}(\ell_1,\ell_2)$ record that $T(\ell_1)=\ell_2$.
\begin{mathpar}
\begin{array}{r@{\,}l}
  \m{tieAuth}(T) \defeq & \own{\authfull{T^{\m{ag}}}} \\[2pt]
  \m{tied}(\ell_1,\ell_2) \defeq & \own{\authfrag{[\ell_1 \mapsto \m{ag}(\ell_2)]}}
\end{array}
\end{mathpar}
where $T^{\m{ag}}$ is the pointwise lifting of $T$ to agreement values.
The use of $\Ag(\m{Loc})$ ensures that
two fragments for the same left location agree on
the right location.

The mapping between locations must be well-formed.
We state this using an assertion $\tieWF(T,h_1,h_2)$,
meaning that $T$ is injective (required for the map to be a bijection),
tied locations occur in their respective heaps,
and both cells are either live or deallocated.
\begin{mathpar}
\begin{array}{l}
  \tieWF(T,h_1,h_2) \defeq \\
  \quad (\all{\ell_1,\ell'_1,\ell_2} T(\ell_1)=\ell_2 \wedge T(\ell'_1)=\ell_2 \implies \ell_1=\ell'_1) \\
  \quad {}\wedge (\all{\ell_1,\ell_2} T(\ell_1)=\ell_2 \implies \ell_1\in\operatorname{dom}(h_1) \wedge \ell_2\in\operatorname{dom}(h_2)) \\
  \quad {}\wedge (\all{\ell_1,\ell_2} T(\ell_1)=\ell_2 \implies
    (\m{live}(h_1,\ell_1) \iff \m{live}(h_2,\ell_2)))
\end{array}
\end{mathpar}

Given a candidate value relation $\Vrel$,
related heaps must also contain related \emph{values} at every pair of tied locations.
We denote this condition by $\heapWF(T,h_1,h_2)$.
\begin{mathpar}
\begin{array}{l}
  \heapWF(T,h_1,h_2) \defeq \all{\ell_1,\ell_2,v_1,v_2} \\
  \quad T(\ell_1)=\ell_2 \wedge h_1(\ell_1)=v_1 \wedge h_2(\ell_2)=v_2
  \implies \later \Vrel(v_1,v_2)
\end{array}
\end{mathpar}
The heap relation $\heapsat(h_1,h_2)$ is then the packaging of these three conditions.
\begin{mathpar}
\begin{array}{l}
  \heapsat(h_1,h_2) \defeq \ex{T} \\
  \quad \m{tieAuth}(T) \sep \tieWF(T,h_1,h_2) \sep \heapWF(T,h_1,h_2)
\end{array}
\end{mathpar}
There are two complementary reasons for the $\later$ in $\heapWF$.
Technically, $\Hrel$ and $\Vrel$ are defined mutually recursively: $\Hrel$ depends on $\Vrel$ through $\heapWF$,
while $\Hrel$ is eventually used to define $\Vrel$.
The $\later$ makes this recursive definition well-founded.
Intuitively, this also makes sense because, if two locations are related,
we must first spend one step reading the locations before we can inspect their contents and determine that the resulting values are related.
Thus, the values obtained from related locations are related later.

$\Hrel$ is preserved by all heap manipulations programs may perform.
\begin{mathpar}
\inferrule*[right=Update]
  {\Hrel(h_1,h_2) \sep \m{tied}(\ell_1,\ell_2) \sep \later \Vrel(v_1,v_2)}
  {\Hrel(h_1[\ell_1\mapsto v_1],h_2[\ell_2\mapsto v_2])}

\inferrule*[right=Free]
  {\Hrel(h_1,h_2) \sep \m{tied}(\ell_1,\ell_2)}
  {\Hrel(h_1[\ell_1\mapsto\none{}],h_2[\ell_2\mapsto\none{}])}

\inferrule*[right=Alloc]
  {\ell_i\notin\operatorname{dom}(h_i)\ (i\in\{1,2\}) \\
   \Hrel(h_1,h_2) \sep \later \Vrel(v_1,v_2)}
  {\bupd\bigl(\Hrel(h_1[\ell_1\mapsto v_1],h_2[\ell_2\mapsto v_2])
    \sep \m{tied}(\ell_1,\ell_2)\bigr)}
\end{mathpar}
Updating tied locations requires the values they are updated with to be related one step later.
We write $h[\ell\mapsto v]$ for an update of $\ell$ to a live cell containing $v$.
Similarly, tied locations can be freed, which replaces both cells with the tombstone $\none{}$.
Only allocation of tied locations changes the ghost state: a basic update extends $T$ and produces a new $\m{tied}$ fragment.

\subsection{The logical relation}
We can now define the relations on contexts and expressions, which are also mutually recursive with $\Vrel$.
They depend on the observation relation $\Orel$, and so need $\Hrel$ to talk about related heaps.
\[
\begin{aligned}
\Krel(K_1,K_2) \defeq {} &
  \all{v_1,v_2} \Vrel(v_1,v_2) \wand \\
& \all{h_1,h_2} \heapsat(h_1,h_2) \wand \\
& \Orel(\econfig{h_1}{K_1[v_1]},\econfig{h_2}{K_2[v_2]})\\[2pt]
\Erel(e_1,e_2) \defeq {} &
  \all{K_1,K_2} \Krel(K_1,K_2) \wand \\
& \all{h_1,h_2} \heapsat(h_1,h_2) \wand \\
& \Orel(\econfig{h_1}{K_1[e_1]},\econfig{h_2}{K_2[e_2]})\\[2pt]
\Rrel(e_1,e_2) \defeq {} &
  \all{K_1,K_2} \later\Krel(K_1,K_2) \wand \\
& \all{h_1,h_2} \heapsat(h_1,h_2) \wand \\
& \Orel(\econfig{h_1}{K_1[e_1]},\econfig{h_2}{K_2[e_2]})
\end{aligned}
\]

$\Rrel$ is $\Erel$ with a $\later$ on $\Krel$.
They differ in that $\Erel$ is the main expression relation and must also relate values,
in which case no step occurs and $K_1[e_1]$ and $K_2[e_2]$ are evaluated \emph{now},
so no later can be afforded.
$\Rrel$ is for expressions guaranteed to take a step, and is used to define $\Vrel$.
\[
\begin{aligned}
  \Vrel(\vlambda{x}{e_1},\vlambda{x}{e_2}) \defeq {}
  & \pers\all{v_1,v_2} \later \Vrel(v_1,v_2) \wand \\
  & \Rrel((\vlambda{x}{e_1})\ap v_1, (\vlambda{x}{e_2})\ap v_2)\\
  \Vrel(\ell_1,\ell_2) \defeq {} & \m{tied}(\ell_1,\ell_2)\\
  \Vrel(v_1,v_2) \defeq {} & \pfalse \quad\qquad\text{otherwise}
\end{aligned}
\]
For lambda abstractions, $\Vrel$ relates those whose applications are in $\Rrel$ when given $\later \Vrel$-related arguments.
This relation must be persistent, since a pair of related lambda abstractions may be used any number of times.
Additionally, the arguments can only be used after one step of $\beta$-reduction, so they only need to be related later.
Locations are related if they are $\m{tied}$. All other mismatched value shapes are unrelated.

\subsection{Open terms}
To deal with open terms, we define a relation $\Grel$ on substitutions: two substitutions are related when they map every variable to $\Vrel$-related values.
$\Erelo$ relates open terms by closing them under related substitutions.
\[
\begin{aligned}
  \Grel(\gamma_1,\gamma_2) \defeq {} & \all{x}\Vrel(\gamma_1(x),\gamma_2(x)) \\
  \Erelo(e_1,e_2) \defeq {} & \all{\gamma_1,\gamma_2}\Grel(\gamma_1,\gamma_2) \wand \Erel(e_1[\gamma_1],e_2[\gamma_2])
\end{aligned}
\]

With the relations defined, we can state and prove the fundamental theorem.
\begin{mathpar}
\m{LocFree}(e)\implies\ptrue \proves \Erel^\circ(e,e)
\end{mathpar}
The $\m{LocFree}(e)$ restriction reflects the fact that the fundamental theorem is stated for source programs.
Source programs cannot contain literal locations, which only arise at runtime.
Moreover, a literal location cannot be interpreted without a pair of heaps, which is not available in the definition above.

\section{Case Study: Iris on Indirection Theory}
\label{sec:iris}

In our second case study, we reconstruct an Iris-style program
logic~\cite{DBLP:journals/jfp/JungKJBBD18} using the base logic.
We cover three main aspects:
invariants,
user-definable ghost state,
and weakest preconditions.
The object language is \emph{HeapLang}, an untyped, call-by-value, concurrent lambda calculus with a heap.
Its small-step operational semantics reduces a \emph{thread pool} (a list of expressions) over a heap.

\subsection{The ghost heap}
The first thing we will need is an abstract heap to model
various assertions about ownership of ghost state. We thus call this heap the \emph{ghost heap},
and define it to be a finite map from ghost names to \emph{cells}.
The type of cells is parameterized by a resource algebra functor $M$.
\[
  \GHeap(M)(P) \defeq \m{Loc} \finmapsto M(P)
\]
$\GHeap(M)$ is itself a unital resource algebra functor.
We can thus instantiate $\Knot$ and $\HPred$ with $\GHeap(M)$; we do so for the remainder of this section.
For convenience when working with $GHeap$, we define ownership of a single ghost cell by specialising $\own{\cdot}$ to a singleton heap. Given a location $\iota$ and a coherent family $a : \all{n} M(\RawPredAt_n)$, the family $\lambda n.\,[\iota\mapsto a_n]$ is also coherent, and we define
\[
  \ownGhost{\iota}{a} \defeq \own{(\lambda n \mapsto [\iota \mapsto a_n])}.
\]
We follow the same convention as for $\own{\cdot}$: we also write $\ownGhost{\iota}{a}$ when $a$ is an element of a resource algebra $A$, assuming that $A$ has been lifted to a resource algebra functor via the constant functor $C$.

It remains to provide a usable type of cells.
Intuitively, it is a sum of a number of resource algebras.
We introduce each resource algebra as it is needed below, before defining their sum.
A more precise definition can be found in the formal development.
Our goal in this section is to provide intuition for how the ghost heap is organised and made extensible, and how it supports the resulting program logic.

\subsection{The concrete heap}
Our model must provide meaning to the points-to assertion $\ell \mapsto v$, which denotes ownership of a piece of the \emph{concrete heap}, the actual heap that programs execute on.
The concrete heap occupies a single cell of the ghost heap, at a distinguished location $\gheap$.
We formulate the concrete heap as a resource algebra $\m{Heap}(V)$ over a set of values $V$.
We define it as a composition of resource algebras.
\[
  \m{Heap}(V) \defeq \m{AuthMap}(\m{Ex}(V))
\]

The authoritative component models the whole concrete heap, while fragments model points-to claims to individual cells.
Since map values are exclusive, two points-to facts cannot refer to the same location.
Concretely, points-to assertions $\ell\mapsto v$ represent ownership of a fragment
\[
  \ell\mapsto v \defeq \ownGhost{\gheap}{(\authfrag{[\ell\mapsto\m{ex}(v)]})}
\]
and the \emph{state interpretation} $\stateinterp(\concheap)$ is a new assertion denoting ownership of the entire concrete heap $\concheap$
\[
    \stateinterp(\concheap) \defeq \ownGhost{\gheap}{(\authfull{\concheap^{\m{ex}}})}
\]
where $\concheap^{\m{ex}}$ is its pointwise lifting to exclusive values.
$\stateinterp$ is used to define the weakest precondition later, but does not normally appear in user-level proofs.

The use of an authoritative map over exclusive values ensures that fragments are consistent with the whole heap.
\[
  \stateinterp(\concheap)\sep\ell\mapsto v
  \proves
  \concheap(\ell)=v
\]
Owning both $\stateinterp(\concheap)\sep(\ell\mapsto v)$ allows updates
\[
  \stateinterp(\concheap)\sep\ell\mapsto v
  \proves
  \bupd\bigl(\stateinterp(\concheap[\ell:=w])\sep\ell\mapsto w\bigr)
\]
and if $\ell\notin\m{dom}(\concheap)$, there is no compatible fragment mentioning $\ell$, allowing allocation.
\[
  \stateinterp(\concheap)
  \proves
  \bupd\bigl(\stateinterp(\concheap[\ell:=v])\sep\ell\mapsto v\bigr)
\]
These rules capture precisely the intuition about concrete heap manipulation in a program,
and they are used to prove weakest precondition rules for heap-manipulating operations.

\subsection{Invariants}
Next, our model gives meaning to \emph{invariants}: named assertions whose resources are maintained throughout execution.
An invariant body $P$ can be any proposition.
In particular, $P$ may own resources from the concrete heap or ghost state.
Invariants are moreover \emph{impredicative}: $P$ may itself mention invariants, including the invariant of which $P$ is the body.

$\knowInv{\iota}{P}$ asserts that the invariant named by $\iota$ has body $P$.
Invariants may be \emph{allocated} from $\later P$, yielding $\knowInv{\iota}{P}$ with a fresh name $\iota$.
They may be \emph{opened} to obtain $\later P$,
and must later be \emph{closed}, relinquishing the resources and reestablishing the invariant.
An invariant cannot be opened while it is already open, for soundness.

To implement this,
ghost heap cells must contain at least three more resource algebras.
First, invariants themselves, where
$\iota$ is a location in the ghost heap storing an approximation of $P$.
Since $P$ is a proposition, we use the resource algebra functor $\m{InvF}$ to map its approximation to a resource algebra.
Second, an invariant \emph{registry}, the set of allocated invariant names, stored at the distinguished location $\gregistry$.
Third, a \emph{mask},
a finite or cofinite\footnote{A cofinite set is one whose complement is finite. The set of all invariant names is cofinite because its complement is empty.}
set of invariant names.
\[
\begin{array}{r@{\,}l}
  \m{InvF}(P) &\defeq \Ag(P) \\
  \m{Reg} &\defeq \m{AuthMap}(\Ag(\m{Unit})) \\
  \m{Mask} &\defeq \DisjMask\footnotemark
\end{array}
\]
\footnotetext{The $\DisjMask$ resource algebra allows splitting/joining these sets disjointly.}

We can now understand the definition of $\knowInv{\iota}{P}$.
It records the agreement that $P$ is the invariant body stored at $\iota$,
as well as ownership of a fragment of the registry indicating that the name $\iota$ has been allocated.
\begin{mathpar}
  \knowInv{\iota}{P} \defeq
    \ownGhost{\iota}{\m{ag}(P)} \sep \ownGhost{\gregistry}{(\authfrag{[\iota\mapsto\m{ag}(\m{unit})]})}
\end{mathpar}
We then define a new assertion $\wsat$ to keep track of which invariants are allocated and govern their allocation, opening, and closing.
It plays an analogous ``global'' role to $\stateinterp$, and is likewise used to define the weakest precondition.
\begin{mathpar}
  \begin{array}{r@{\,}l}
  \wsat \defeq &
    \ex{S : \m{Set}(\m{Loc})} \regAuth(S) \\
    & \sep \bigsep_{\iota\in S}\slot(\iota) \sep \disabled(\top\setminus S) \\
  \regAuth(S) \defeq & \ownGhost{\gregistry}
    {(\authfull{[\iota\mapsto\m{ag}(\m{unit})]_{\iota\in S}})}
  \end{array}
\end{mathpar}
$\wsat$ owns the registry, a finite set of allocated invariant names $S$.
$\regAuth$ asserts exclusive ownership of the registry, similar to $\stateinterp$.

Each invariant $\iota\in S$ is in one of two states.
\[
\begin{array}{r@{\,}l}
  \slot(\iota) \defeq
  & \bigl(\ex{P} \ownGhost{\iota}{\m{ag}(P)} \sep \later P \sep \disabled(\{\iota\})\bigr) \\
  & \vee\ \enabled(\{\iota\})
\end{array}
\]
The first disjunct represents a closed invariant: $\wsat$ owns its body (under $\later$) together with a $\disabled$ token,
which may be understood as the obligation to close it.
The second represents an open invariant, where the body and $\disabled$ token are absent,
and $\wsat$ has an $\enabled$ token, indicating that the invariant is currently open.
The protocol allows these tokens and resources to move in and out of $\wsat$.
An invariant cannot be opened while it is already open, since two copies of $\enabled$ cannot be owned for the same mask.
Both tokens are defined for masks $E$ using ownership of disjoint masks, and require two additional dedicated ghost names, $\gname_{en}$ and $\gname_{dis}$.
\[
  \begin{aligned}
    \enabled(E) &\defeq \ownGhost{\genabled}{\disjmask(E)} \\
    \disabled(E) &\defeq \ownGhost{\gdisabled}{\disjmask(E)}
  \end{aligned}
\]

\subsection{User-defined ghost state}
Having seen the setup of the concrete heap and invariants,
we can now define the cell type mentioned when we first talked about the ghost heap.

As a first approximation, we define it as the sum of the four resource algebras used by the program logic.
$\m{InvF}$ depends on the type $P$ of propositions stored in a cell, so it is a functor.
We turn the others into functors using the constant resource algebra functor $C$, and define:
\[
  \basecell(V) \defeq
  C(\m{Heap}(V)) + \m{InvF} + C(\m{Reg}) + C(\m{Mask})
\]
$(\cdot + \cdot)$, a sum of resource algebra functors, is a resource algebra functor.
Thus $\basecell(V)$ is also a resource algebra functor.

This would be sufficient if we only ever worked with the aforementioned kinds of ghost state.
However, the motivation for having a rich vocabulary of resource algebras is that users can compose them, to build new problem-specific abstractions, like transition systems or protocols.
These would have to be stored in the ghost heap as well.
As such, the cell type of the ghost heap should be somehow left extensible.

We do this by following the recipe given in \emph{Data types \`a la
carte}~\cite{DBLP:journals/jfp/Swierstra08}.
To illustrate this,
suppose the user wants to extend the ghost heap with a proposition-free resource algebra $A$.
A natural first step would be for the user to replace $\basecell(V)$ with
\[
  M \defeq C(A) + \basecell(V)
\]
Assuming the sum is nested to the right,
the injection of the invariant resource into $\basecell(V)$ is
\[
  \begin{aligned}
    p : \m{InvF}(P) & \longmapsto \m{inr}(\m{inl}(p)) : \basecell(V)(P)
  \end{aligned}
\]
whereas the corresponding injections into $M$ are
\[
  \begin{aligned}
    a : A & \longmapsto \m{inl}(a) : M(P) \\
    p : \m{InvF}(P) & \longmapsto \m{inr}(\m{inr}(\m{inl}(p))) : M(P)
  \end{aligned}
\]
The path for $\m{InvF}$ changed by one $\m{inr}$ constructor.
This would be a problem if existing client code asserted ownership of an invariant by mentioning the path concretely, as it would then have to be changed to accommodate the extension.
The same issue would occur for $A$'s path if we further extended $M$ with another sum.

A better approach would be to abstract over the concrete path by synthesising an injection
$\m{InvF}(P) \to M(P)$ and proofs of associated laws.
Hence,
we use a typeclass $\m{InG}(M,F)$, with one main operation $\m{inj}_P : F(P) \longrightarrow M(P)$,
which must be shown to respect the resource algebra structure.
The proofs are assembled compositionally by typeclass search.

Clients then only need $\m{InG}$ instances to assert that the functors they need are present in an ambient cell type $M$, rather than requiring $M$ to be exactly $\basecell(V)$ or some other concrete type.
Consequently, for the resource algebra $A$ from earlier, the ownership assertion $\ownGhost{\gamma}{a}$ requires only the membership $\m{InG}(M,C(A))$.
The added component may more generally be a resource algebra functor $F$, allowing user-defined ghost state to contain propositions.

This is one place where our development diverges from Iris, which uses (conceptually) a list of ghost heaps, making ghost names tuples of a list index and a heap location, instead of an extensible sum.
We take this route largely for simplicity.

\subsection{Weakest preconditions}
Following Iris, the weakest precondition is the assertion
\[
\begin{array}{@{}l@{}}
  \wpE{\mask}{e}{\Phi} \defeq
  \all{\sigma}\wpstate(\sigma) \wand \bupd\exZero \\
  \quad \begin{array}{@{}l@{}}
    \text{if } e=\m{Val}(v), \text{ then } \wpstate(\sigma)\sep\Phi(v); \\
    \text{if } e\notin\m{Val}, \text{ then } \exZero\m{reducible}(e,\sigma) \land \\
    \begin{array}{@{}l@{}}
      \all{(e',\sigma',\vec e_f)\in\m{step}(e,\sigma)}
        \bupd\later\bupd\exZero
        \bigl(\wpstate(\sigma') \sep\wpE{\mask}{e'}{\Phi} \\
          \quad\sep\bigsep_{e_f\in\vec e_f}\wpE{\top}{e_f}{(\lambda\_.\ptrue)}
        \bigr)
    \end{array}
  \end{array}
\end{array}
\]
The resources threaded through the weakest precondition are given by
$\wpstate(\sigma) \defeq \stateinterp(\sigma)\sep\wsat\sep \enabled(\mask)$.
If $e$ is a value, the postcondition $\Phi$ must hold immediately,
while the resources in $\wpstate(\sigma)$ are preserved.
Otherwise, $e$ must be reducible, and for every possible new configuration $(e',\sigma')$,
$e'$ must satisfy the weakest precondition $\wpE{\mask}{e'}{\Phi}$ with $\wpstate(\sigma')$ preserved.
The list $\vec e_f$ contains the threads forked during a reduction step.
Each forked thread must satisfy a weakest precondition with postcondition $\ptrue$,
meaning that it must execute safely. Its return value is discarded.

The argument $\mask$ is a mask specifying the invariants that may be opened during the execution of the expression.
It must be restored after each physical step.
Forked threads begin with no invariants open and therefore use the full mask $\topmask$.
The $\exZero$s waive the obligations in the reducible case at index 0,
where there are no steps remaining and we cannot distinguish stuck expressions from reducible ones.
The $\bupd\later\bupd$ allows ghost state updates both before and after a logical step.
A common example is proving that an invariant is preserved by an atomic step:
the first basic update opens the invariant and obtains its body under $\later$,
the physical step consumes the $\later$,
and the second update can then reestablish and close the invariant.
The $\later$ both synchronises program steps with logical steps and guards the recursion.

We may then define the Hoare triple as
\[
  \hoare{P}{e}{\Phi}{\mask} \defeq \pers\bigl(P \wand \wpE{\mask}{e}{\Phi}\bigr)
\]
and use it to derive all the standard program logic rules.

\subsection{Verified programs}

We validate the program logic on simple but characteristic examples.
First, parallel increment, where two threads
perform a non-atomic increment on a shared location.
We prove specifications of varying strength using different techniques, including a plain invariant, and a three-state protocol using a combination of the one-shot and fraction resource algebras.

Second, we verify Landin's Knot~\cite{DBLP:journals/cj/Landin64}:
both by ordinary induction and by L\"ob induction~\cite{birkedal2020lecture}, where an invariant owns the self-referential cell.
The resulting proof exercises impredicative invariants, persistent higher-order specifications, and guarded recursion together.

\subsection{Discussion}

This case study validates the claim that \ind{} can serve as an alternative foundation for Iris.
To this end, the construction of the program logic intentionally follows Iris closely, to sufficient depth that we can reconstruct a program logic usable for verifying simple programs.

The proofs are written using Lean tactics for rewriting logical entailments, permuting separating conjunctions, frame inference, biabduction, and ramified Hoare triples.
They therefore resemble conventional~\cite{slf} interactive separation-logic proofs;
an equivalent to the Iris Proof Mode~\cite{DBLP:journals/pacmpl/KrebbersJ0TKTCD18} is not implemented.
Reuse of the Iris-Lean implementation of it~\cite{irislean} is likely to be possible, given that our separation logic is a standard affine BI logic.

\section{Related Work}
\label{sec:related}

\paragraph{From step-indexing to higher-order separation logic}
Step-indexing originated as a technique for modelling
recursive types~\cite{DBLP:journals/toplas/AppelM01}
and higher-order state~\cite{DBLP:conf/lics/AhmedAV02}.
\textcite{DBLP:conf/popl/AppelMRV07} then abstracted it into the $\later$ operator of a modal logic,
a presentation that later works~\cite{DBLP:conf/lics/DreyerAB09}
established as a standard way of handling step-indexing.
\textcite{DBLP:conf/esop/HoborAN08} used this approach to build a higher-order concurrent
separation logic, later generalising the construction to \ind~\cite{DBLP:conf/popl/HoborDA10,msl}.
Meanwhile, \citet{DBLP:journals/tcs/BirkedalST10} developed a different construction,
based on ultrametric spaces, and used the theorem of \citet{DBLP:journals/jcss/AmericaR89} to solve
recursive domain equations.
This construction, adapted to (complete) ordered families of equivalences and (unital) cameras,
is the foundation of Iris~\cite{DBLP:journals/jfp/JungKJBBD18}.

\paragraph{\ind}
\ind{} was introduced by \citet{DBLP:conf/popl/HoborDA10} and
mechanised in Rocq in the Mechanized Semantic Library (MSL)~\cite{msl}.
MSL was later extended and used as the foundation of VST~\cite{DBLP:conf/nfm/Appel12}.
We mostly follow the MSL development, with the main difference being that we
replace separation algebras with Iris-style resource algebras.
Additionally, MSL supports some minor generalisations beyond our presentation.
For example, MSL supports functors of mixed variance, whereas ours are covariant.
These extensions are straightforward and not central to our development,
so we omit them for ease of implementation and presentation.

\newcommand*{\fstar}{F$\star$}

\paragraph{PulseCore}
The closest system to ours is PulseCore~\cite{ebner2025pulsecore}, an impredicative concurrent separation logic embedded in \fstar.
It also uses \ind{} as a foundation for reasons of simplicity,
and arrives at a very similar ``base logic'' of hereditary predicates, with two main differences.

The first is in the construction of the knot.
PulseCore recursively defines the resources rather than the propositions,
and a stored proposition is a dependent function over knots of lower level rather than a sequence of approximations.
They report that \fstar{}'s extensional type theory makes this a shorter construction;
the advantage does not carry over to Lean's intensional type theory,
so we follow the classic approach of defining propositions recursively.

A second difference is in what the knot is instantiated with.
PulseCore fixes the higher-order ghost state to
a heap whose cells hold only either an invariant or a stored predicate.
First-order ghost state is contained in a separate heap
and other higher-order ghost state is obtained by encoding.
Its authors note that they could ``make the construction parametric in the supported
higher-order ghost state cells and their associated functors'',
but decided against ``the API complexity of extra parameters''.
This is the generalisation we make,
in parameterising the knot by an arbitrary resource algebra functor,
so higher-order ghost state is user-definable, like in Iris.

Other practical differences in PulseCore outside of its model construction are that
it is meant for intrinsic proofs about F$\star$ programs, so the step index is decreased by ghost code
rather than by steps of the reduction relation,
and a L\"ob induction principle is not exposed at all,
since \fstar's native well-founded recursion suffices.

\paragraph{VST}
VST~\cite{DBLP:conf/nfm/Appel12} is a program logic for C, and the most substantial system built on \ind.
VST's resource model is closely tied to CompCert~\cite{Leroy-BKSPF-2016}'s memory model,
so that the program logic can be proven sound against CompCert's operational semantics.
Verification in VST then combines with CompCert's compiler correctness to give guarantees
about compiled code. In particular, VST's resource is a heap, in which a location holds
either a value with a permission, or a predicate, such as a function's specification or a
lock's invariant.

\textcite{DBLP:journals/corr/abs-2207-06574} brought some of Iris's ghost state into VST to increase its
expressiveness, while keeping its resource model tied to CompCert. Ghost state was added as a second
component of the resource
\[
  \m{heap} \;\times\; \m{ghost}
\]
and invariants were derived from it in the same work. However, this approach still could not fully support
higher-order ghost state, due to universe inconsistencies.

This limitation motivated \textcite{DBLP:journals/pacmpl/ManskyD24} to rebuild VST's program logic on top
of Iris, from which higher-order ghost state is then inherited. Both works left open the question of
whether \ind{} can be made as expressive as Iris. Our work addresses this question.

\paragraph{Iris and Iris-Lean}
Iris~\cite{DBLP:journals/jfp/JungKJBBD18} is the frontier of higher-order concurrent separation logic,
with a large body of features and applications built on it, and is mechanised in Rocq.
Iris-Lean~\cite{konig2022improved,irislean} is a faithful port of Iris to Lean,
and, to our knowledge, the only other mechanised step-indexed logic in Lean.
It follows Iris's construction directly, up to minor differences between the two proof assistants,
so the comparison to Iris's base logic in
\cref{sec:background,sec:library}
applies to it unchanged.

\section{Conclusion}
\label{sec:conclusion}

Our aim in this work was to clarify and generalise the foundations of a higher-order separation logic, thereby making the construction of higher-order semantic models and program logics accessible to a wider audience, and spurring the development of more expressive verification tools.
To this end, we presented a Lean implementation of \ind{}, generalised to support Iris-style higher-order ghost state.
Its construction is elementary and uses only dependent types, without domain theory or category theory.
It should thus serve as a recipe for implementing similar models.
We demonstrated the expressiveness of the resulting separation logic for constructing both logical relations and an Iris-style program logic,
and also clarified the relationship between \ind{} and Iris's model.

Some parts of Iris remain to be reconstructed.
The most immediate direction for future work is to reconstruct the recent work on discharging $\later$ on these foundations: later credits~\cite{DBLP:journals/pacmpl/SpiesGTJKBD22}, and the physical-step modality~\cite{DBLP:conf/cpp/0001HGK26}.

\printbibliography

\end{document}